# Superhumps and post-outburst rebrightening episodes in the AM CVn star SDSS J012940.05+384210.4

Jeremy Shears, Steve Brady, Robert Koff, William Goff & David Boyd

## Abstract

We report unfiltered photometry of the first confirmed outburst of the AM CVn system SDSS J012940.05+384210.4 during 2009 December. At its brightest the star was magnitude 14.5, 5.4 magnitudes above mean quiescence. Although the first part of the outburst was not observed, six remarkable rebrightening events were observed during the course of the outburst. Forty-one days after the outburst was detected, the star was still 1.7 magnitudes above quiescence. Superhumps were observed during the outburst with a peak-to-peak amplitude of 0.06 mag and $P_{sh}$ = 37.9(2) min. We also used archival data to show that another AM CVn system, SDSS J124058.03-015919.2, has also undergone at least one outburst, with an amplitude of ~4.6 magnitudes.

## Introduction

The AM CVn stars are a class of cataclysmic variables comprising two interacting white dwarfs. Due to their proximity, mass is transferred from one star (the secondary, or donor) and because the material has significant angular momentum it does not settle immediately on the primary, but instead forms an accretion disc around the primary. Spectroscopy has shown that these binaries are comprised largely, or completely, of helium. They have ultra-short orbital periods in the range ~10 to 65 mins. These systems are so compact that both binary components are degenerate (or semi-degenerate) and it is thought that the mass transfer is driven by gravitational radiation (1) (2) .

Very few confirmed AM CVn stars are known: only about two dozen are suspected at the time of writing (3). Anderson *et al.* (2) examined spectra from the Sloan Digital Sky Survey (4) (5) (SDSS) in an attempt to find new AM CVn stars from their telltale helium-dominated optical spectrum. SDSS J012940.05+384210.4 was identified in this search from its broad HeI emission lines (2). Thus far, no orbital period has been published. SDSS lists magnitude g=19.81, r=20.04. Data from the Catalina Real-Time Sky Survey (CRTS) (6), shows that it varies between V= 18.9 and 21.1, with a mean of V= 19.9. The star is located in Andromeda at RA 01h 29min 40.05s +38deg 42min 10.4s.

The outburst of SDSS J012940.05+384210.4 discussed in this paper, which is the first on record, was discovered on 2009 Nov 29.045 by the authors (7) using the

Bradford Robotic Telescope (8) at an unfiltered magnitude of 14.5 as part of a programme to monitor cataclysmic variables for outbursts using small telescopes; this particular star has been monitored since 2009 January.

**Photometry and analysis**

The authors conducted 63 h of unfiltered photometry using the instrumentation shown in Table 1 and according to the observation log in Table 2. Images were dark-subtracted and flat-fielded prior to being measured using differential aperture photometry relative to the V-sequence given on AAVSO chart 3325juo (9). Heliocentric corrections were applied to all data.

**Profile of the 2009 outburst**

The overall light curve of the outburst is shown in Figure 1, based on the authors' photometry, supplemented with data from the AAVSO International Database (10), and CRTS. Expanded plots of some of the longer photometry runs are shown in Figure 2.

Unfortunately the beginning of the outburst is not well constrained due to gaps in the observational data and the plateau phase, assuming one was present, was missed. Only a single observation was made on discovery night (JD 2455164), when the star was at its brightest at magnitude 14.5; the previous recorded observation was made 12 days earlier when the object was fainter than magnitude 16.9. Two nights after the outburst was detected the star had faded to 14.8 and was undergoing a rapid decline at 2.7 mag/d (Figure 2a).This was followed by at least 6 rebrightening episodes at 3 to 5 day intervals (~ JD 2455169, 172, 176, 183, 188 and 203). Other rebrightenings might have been missed due to incomplete coverage. Based on an average quiescence brightness V=19.9, the outburst amplitude was therefore at least 5.4 magnitudes. The star was still ~1.7 magnitudes above mean quiescence 41 days after the outburst was first detected.

**Detection of superhumps**

Inspection of the light curve from JD 2455172, during one of the rebrightening episodes, clearly shows the presence of regular modulations which we interpret as superhumps (Figure 2 b). We performed a period analysis of the combined data using the Lomb-Scargle algorithm in *Peranso* (11), after subtracting the mean magnitude. This gave the power spectrum in Figure 3. The highest peak in the power spectrum is at 38.02(22) cycles/day, which we interpret as being due to the superhumps. This results in a superhump period, $P_{sh}$ = 37.9(2) min or 2274(12) sec. The error estimates were derived using the Schwarzenberg-Czerny method (12). The same signal was identified using several other algorithms in *Peranso*. Pre-whitening the power spectrum with the 38.02 cycles/day signal left only weak signals, none of which has any significant relationship to the superhump period. A phase diagram of the data, folded on the proposed $P_{sh}$, is shown in Figure 4, where two cycles are shown for clarity. This shows a classical saw-toothed superhump profile with a peak-to-peak amplitude of 0.06 mag.

We also performed a Lomb-Scargle period analysis on the data from the rebrightening episodes of JD 2455178 (Figure 2c) and 183 (Figure 2d). In both cases the star was fading rapidly (3.1 and 2.3 mag/d respectively), so we first subtracted the linear trend from the data and found signals at 38.16(91) and 37.94(90) cycles/day respectively. These values are consistent with the value for $P_{sh}$ measured on JD 2455172 as described above. By contrast, when we analysed the data from JD 2455166, corresponding to the first observed rapid fade (Figure 2a), we only found very weak signals, none of which appeared to be related to the signals at around 38 cycles/day which were found later in the outburst (however, we note that Kato *et al* (13)*.* report a signal in data from JD 2455166 at around 26 mins which they interpret as $P_{sh}$). Unfortunately our time series data from the other nights were too noisy, due to the faintness of the star, for satisfactory period analysis to be carried out.

**Identification of a further outbursting AM CVn system: SDSS J124058.03-015919.2**

We note the similarity of the probable orbital period of SDSS J012940.05+384210.4 (~37 min) to that of SDSS J124058.03-015919.2 (37.355(2) min) (14). The latter was identified as an AM CVn system by Roelofs *et al.* (14) with a low mass transfer rate, although they could not rule out outbursts. We examined CTRS data (6) and found that in quiescence SDSS J124058.03-015919.2 varies between V= 19.1 and 21.4, with a mean of V= 19.9. The star appears to have undergone an outburst in 2005: ASAS-3 (15) (All Sky Automated Survey) detected the outburst on 2005 March 15 and March 17 at V=13.53(04) and 13.78(5) respectively. CRTS also appears to have caught the tail of this outburst at V = 17.03(4) on 2005 April 4 and it was still above quiescence (V=18.3) some 77 days after detection. The outburst light curve is shown in Figure 5.The outburst amplitude was 4.6 magnitudes above mean quiescence. Thus SDSS J124058.03-015919.2 is the seventh confirmed outbursting AM CVn system.

**Discussion**

AM CVn systems have been classified into 3 broad groups, based on their observed characteristics, as a function of increasing orbital periods (1) (14):

(i) The short-period, high state systems which are permanently bright and have high mass transfer rates, with $P_{orb} \lesssim$ 20 min, such as AM CVn and HP Lib

(ii) The longest period, low state, systems that are believed to be in a stable state of low mass transfer, with $P_{orb} \gtrsim$ 40 mins, such as GP Com and CE315

(iii) The intermediate period outbursting systems, with $20 \lesssim P_{orb} \lesssim 40$ mins. Thus far only 5 such systems have been confirmed (16): CR Boo, KL Dra, V803 Cen, CP Eri and V406 Hya (previously known as 2003aw).

The stars in group (iii) have exhibited superhumps, where the superhump period excess, $\varepsilon = (P_{sh} - P_{orb})/P_{orb}$, is in the range 0.003 to 0.020 (see Table 3). Therefore our measurement of $P_{sh} = 37.9(2)$ min for SDSS J012940.05+384210.4 suggests that its $P_{orb}$ is probably ~37 min. The outbursting nature of SDSS J012940.05+384210.4 which we report in this paper clearly makes this system the sixth member of the group (iii) described above.

The group (iii) AM CVn systems have been described as the helium cousins of the hydrogen-dominated dwarf novae (16). Patterson *et al*. (17) and Gaensicke *et al*. (18) analysed $P_{orb}$ as a function of $P_{sh}$ for a large sample of SU UMa hydrogen dwarf novae and found a strong correlation between the two parameters; unfortunately such a correlation does not presently exist for superhumping AM CVn systems as so few have been observed. It is hoped that as more surveys are conducted, more such systems will be identified.

One of the most remarkable features of the outburst light curve of SDSS J012940.05+384210.4 is the rebrightening events. A very similar behaviour was observed by Kato et al. (19) in the AM CVn system, V803 Cen, during its 2003 June outburst, when at least 7 rebrightening events were observed. Rebrightenings (sometimes called "echo outbursts") have also been reported in several members of the WZ Sge family of hydrogen dwarf novae, including WZ Sge itself (20). WZ Sge systems are a highly evolved sub-set of the SU UMa family, which have very short orbital periods and low mass transfer rates.

**Conclusions**

The AM CVn system SDSS J012940.05+384210.4 was found to be in outburst in 2009 December, the first confirmed outburst on record. We report unfiltered photometry obtained during the outburst. At its brightest the star was magnitude 14.5, 5.4 magnitudes above the mean quiescence level of V = 19.9 found in data from the Catalina Real-Time Sky Survey (6). Although the first part of the outburst was not observed, six remarkable rebrightening events were observed during the course of the outburst. The star was still 1.7 magnitudes above quiescence 41 days after the outburst was detected. Superhumps were observed during the outburst, confirming it to be a superoutburst, with a peak-to-peak amplitude of 0.06 mag and $P_{sh} = 37.9(2)$ min.

SDSS J012940.05+384210.4 is thus confirmed to be a sixth member of the outbursting group of AM CVn stars. The observed rebrightening episodes are similar to those seen during the 2003 June outburst of the AM CVn star, V803 Cen. Moreover, they are reminiscent of the echo outbursts shown by many members of the WZ Sge family of hydrogen dwarf novae.

We also used archival data to show that another AM CVn system, SDSS J124058.03-015919.2, and which likely has a similar $P_{orb}$ to SDSS J012940.05+384210.4, has also undergone at least one outburst, in 2005 March. The outburst amplitude was at least 4.6 magnitudes.

Rather few AM CVn stars are known and the characteristics of the outbursting systems are not well studied. Thus we encourage further monitoring of SDSS J012940.05+384210.4 and SDSS J124058.03-015919.2 with the aim of identifying and studying future outbursts.

## Acknowledgements

The authors acknowledge the use of data from the Catalina Real-Time Transient Survey (CRTS), the All Sky Automated Survey (ASAS-3) and from the AAVSO International Database contributed by Eddy Muyllaert and Ian Miller. We thank Prof. Joe Patterson and Jonathan Kemp, of the Center for Backyard Astrophysics (CBA), for allowing us to use CBA data. We gratefully acknowledge the use of the Bradford Robotic Telescope operated by the Department of Cybernetics, University of Bradford, located on Tenerife and which was used by JS to detect the outburst discussed in this paper. This research made use of SIMBAD and Vizier, operated through the Centre de Donées Astronomiques (Strasbourg, France), and the NASA/Smithsonian Astrophysics Data System. Finally, we would like to thank our referees, Prof. Boris Gänsicke (University of Warwick, UK) and Dr. Chris Lloyd (Open University, UK), for their helpful comments which have improved the paper.

**Addresses**:

JS: "Pemberton", School Lane, Bunbury, Tarporley, Cheshire, CW6 9NR, UK [bunburyobservatory@hotmail.com]

SB: 5 Melba Drive, Hudson, NH 03051, USA [sbrady10@verizon.net]

RK: 980 Antelope Drive West, Bennett, CO 80102, USA [bob@antelopehillsobservatory.org]

WG: 13508 Monitor Lane, Sutter Creek, CA 95685, USA [b-goff@sbcglobal.net]

DB: 5 Silver Lane, West Challow, Wantage, Oxon, OX12 9TX, UK [drsboyd@dsl.pipex.com]

| Observer | Telescope | CCD |
|---|---|---|
| Shears | 0.28 m SCT | Starlight Xpress SXVF-H9 |
| Brady | 0.4 m reflector | SBIG ST-8XME |
| Koff | 0.25 m SCT | Apogee AP47 |
| Goff | 0.4 m reflector | SBIG ST-8 |
| Boyd | 0.35 m SCT | Starlight Xpress SXV-H9 |

**Table 1: Instrumentation used**

| Start time JD | End time JD | Duration (h) | Observer |
|---|---|---|---|
| 2455166.304 | 2455166.463 | 3.8 | Shears |
| 2455166.416 | 2455166.489 | 1.8 | Boyd |
| 2455166.514 | 2455166.881 | 8.8 | Koff |
| 2455166.615 | 2455166.886 | 6.5 | Goff |
| 2455172.214 | 2455172.328 | 2.7 | Shears |
| 2455172.441 | 2455172.737 | 7.1 | Brady |
| 2455176.493 | 2455176.772 | 6.7 | Brady |
| 2455178.449 | 2455178.717 | 6.4 | Brady |
| 2455183.479 | 2455183.705 | 5.4 | Brady |
| 2455184.529 | 2455184.702 | 4.2 | Brady |
| 2455188.485 | 2455188.691 | 4.9 | Brady |
| 2455203.478 | 2455203.495 | 0.4 | Brady |
| 2455206.462 | 2455206.644 | 4.4 | Brady |

**Table 2: Log of time series observations**

| Name | $P_{orb}$ (s) | (min) | $P_{sh}$ (s) | (min) | $\varepsilon$ |
|---|---|---|---|---|---|
| CR Boo | 1471 | 24.5 | 1487 | 24.8 | 0.011 |
| KL Dra | 1500 | 25.0 | 1530 | 25.5 | 0.020 |
| V803 Cen | 1612 | 26.9 | 1618 | 27.0 | 0.003 |
| CP Eri | 1701 | 28.4 | 1716 | 28.6 | 0.009 |
| 2003aw | 2028 (21) | 33.8 | 2042 | 34.0 | 0.007 |

**Table 3: Orbital and superhump periods of outbursting AM CVn systems**

Data are from Nelemans (1) unless otherwise stated

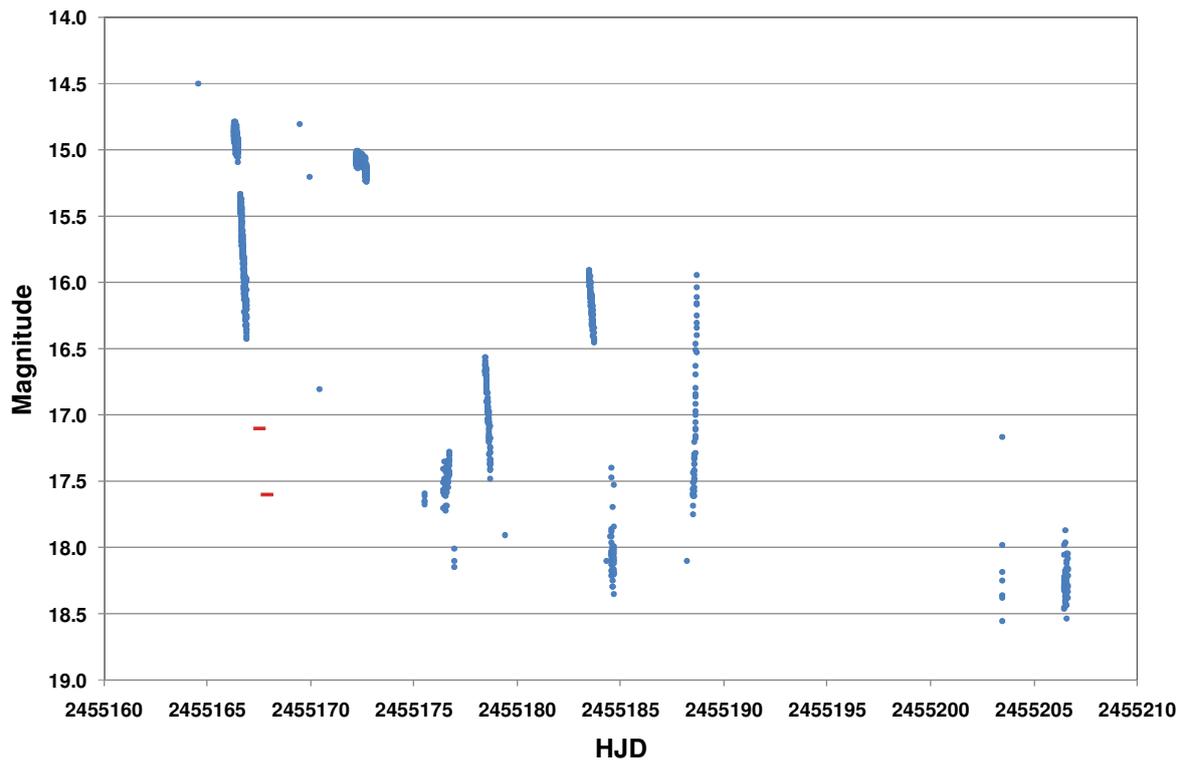

**Figure 1: Outburst light curve of SDSS J012940.05+384210.4**

The red dashes indicate "fainter than" observations

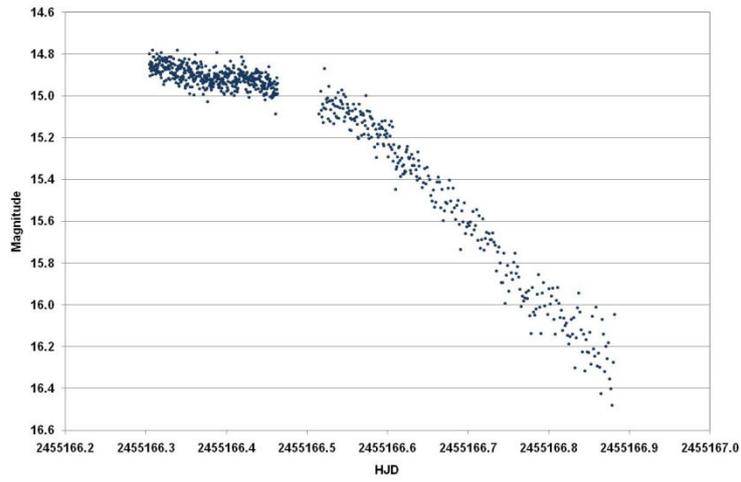

(a) JD 2455166

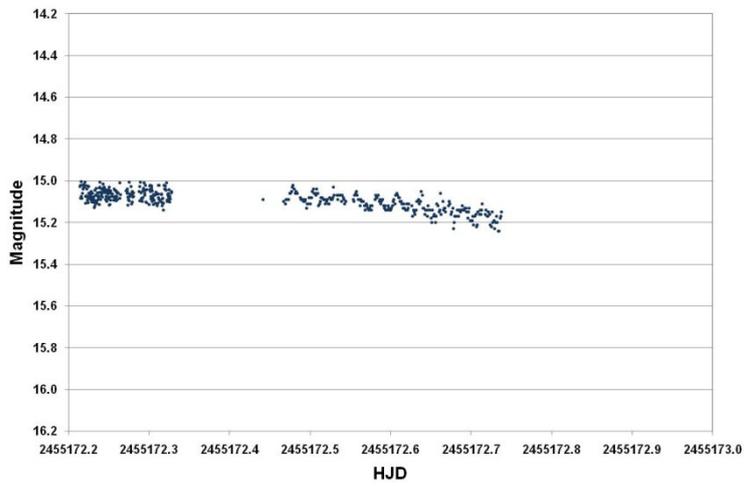

(b) JD 2455172

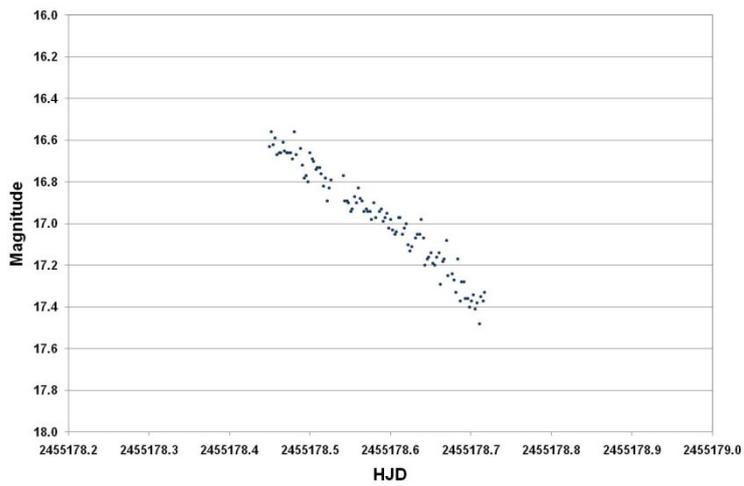

(c) JD 2455178

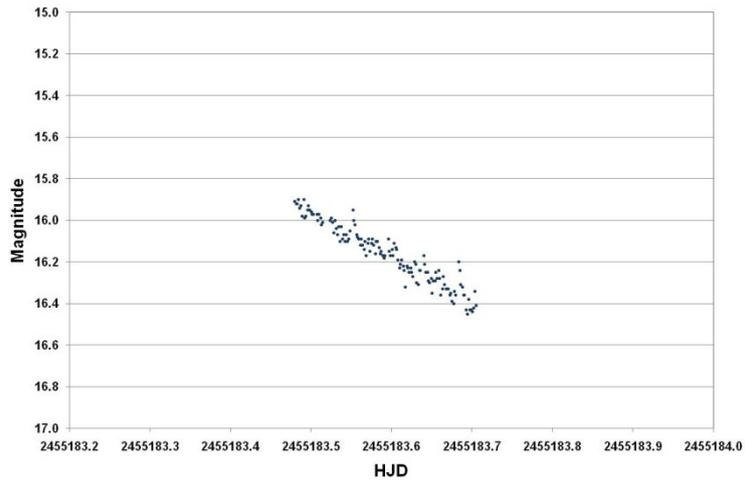

(d) JD 2455183

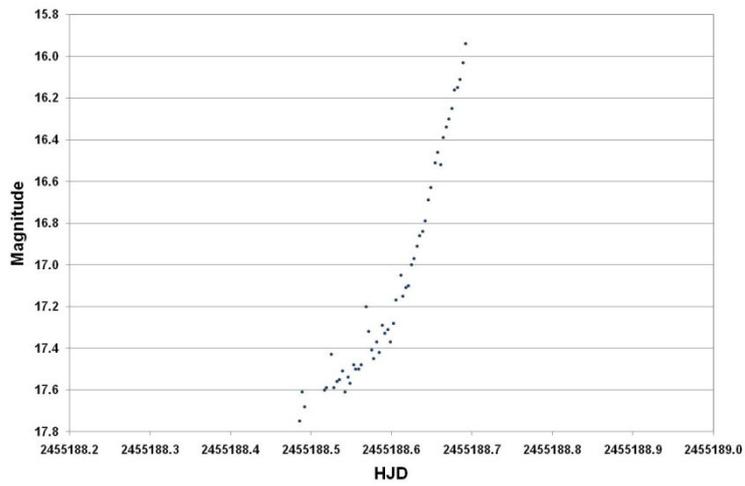

(e) JD 2455188

**Figure 2: Light curves from time resolved photometry**

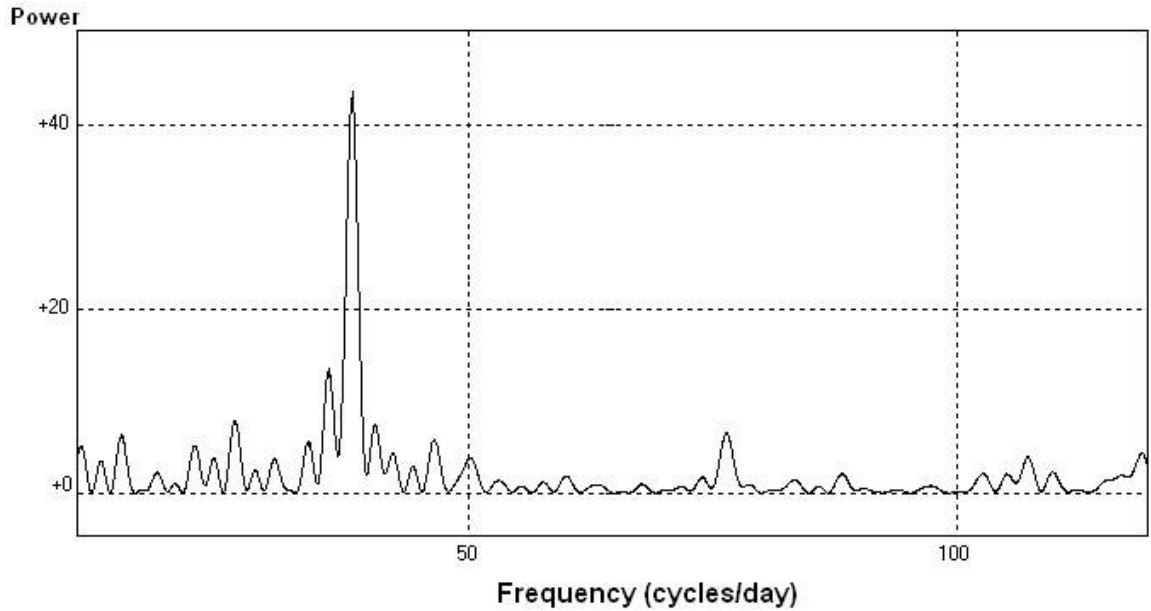

**Figure 3: Power spectrum of data from JD 2455172**

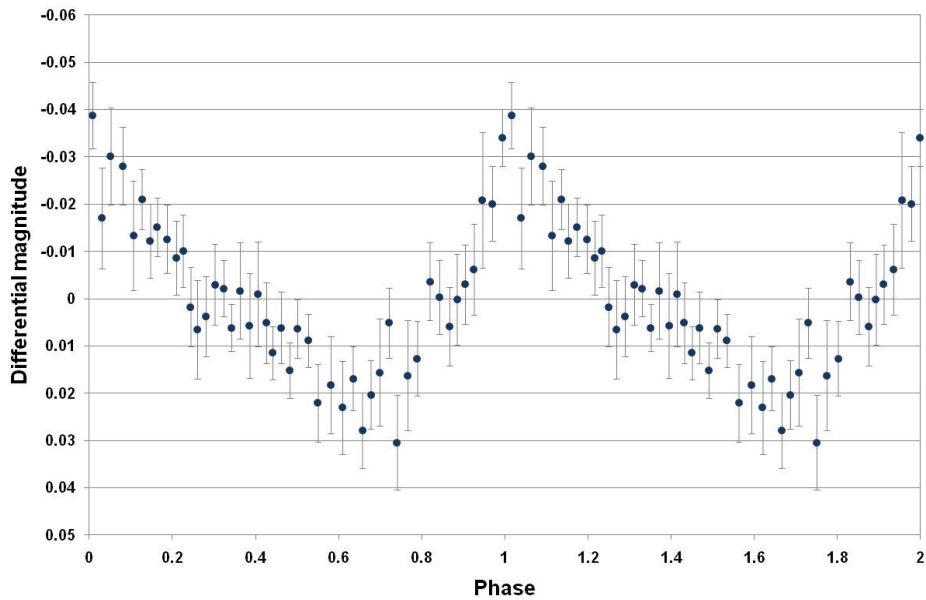

**Figure 4: Phase diagram of data from JD 2455172 folded on $P_{sh}$ = 37.9 min**

Each data point is the mean of 10 individual measurements

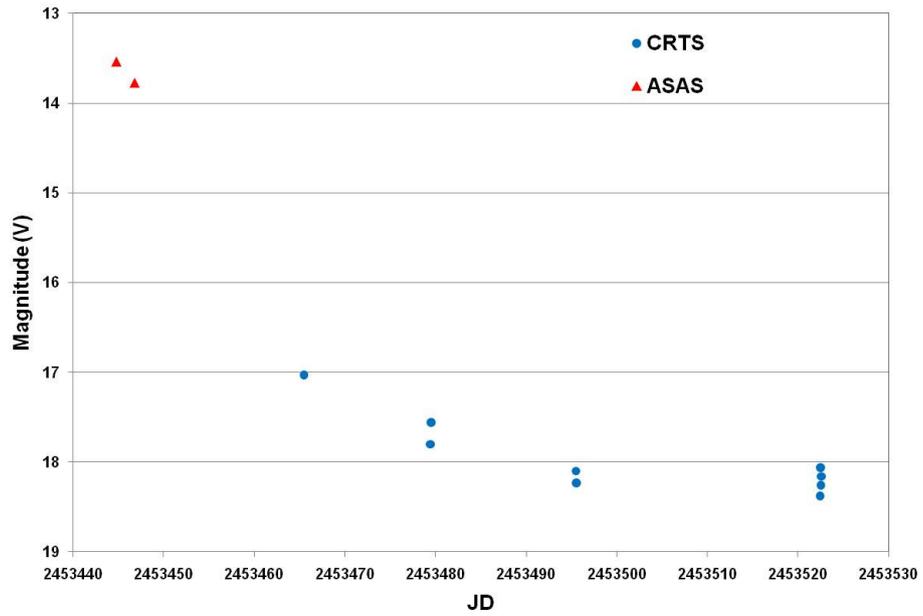

**Figure 5: The 2005 outburst of the AM CVn system SDSS J124058.03-015919.2**

The first data point is on 2005 March 15. Data are from ASAS-3 and CRTS are as indicated